\documentclass[%
 reprint,
superscriptaddress,
 amsmath,amssymb,
 aps,pre,hyphens,floatfix
]{revtex4-2}

\usepackage[]{amsmath}
\usepackage{mathrsfs}
\usepackage{mathtools} 
\usepackage{enumerate}
\usepackage[usenames,dvipsnames,svgnames,table]{xcolor}
\usepackage{IEEEtrantools}
\usepackage{graphicx}
\usepackage{array}
\usepackage{multirow}
\usepackage{verbatim}
\usepackage{soul}
\usepackage{tabularx}
\usepackage{booktabs,siunitx}
\usepackage[T1]{fontenc}

\newcommand{\angs}[1]{\langle #1 \rangle}
\newcommand{\rR}{\mathscr{R}}
\newcommand{\rI}{\mathscr{I}}
\newcommand{\sR}{\mathcal{R}}
\newcommand{\sI}{\mathcal{I}}
\newcommand{\ha}{\hat{a}}
\newcommand{\hb}{\hat{b}}
\newcommand{\trI}{\tilde{\mathscr{I}}}
\newcommand{\trR}{\tilde{\mathscr{R}}}
\newcommand{\tI}{\tilde{\mathcal{I}}}
\newcommand{\tR}{\tilde{\mathcal{R}}}
\newcommand{\ud}{\mathrm{d}}

\newcommand{\mycomment}[1]{}

\begin{document}

\preprint{APS/123-QED}

\title{Critical behavior of the diffusive susceptible-infected-recovered model}

\author{Shengfeng Deng}
\email[]{shengfeng.deng@ek-cer.hu}
\affiliation{Institute of Technical Physics and Materials Science, Center for Energy Research, P.O. Box 49, H-1525 Budapest, Hungary}
\author{G\'eza \'Odor}
\email[]{odor@mfa.kfki.hu}
\affiliation{Institute of Technical Physics and Materials Science, Center for Energy Research, P.O. Box 49, H-1525 Budapest, Hungary}

\date{\today}

\begin{abstract}
The critical behavior of the non-diffusive 
susceptible-infected-recovered model on lattices had been well
established in virtue of its duality symmetry. By performing 
simulations and scaling analyses for the diffusive variant on
the two-dimensional lattice, we show that diffusion for all 
agents, while rendering this symmetry destroyed, constitutes a singular 
perturbation that induces asymptotically distinct dynamical and stationary 
critical behavior from the non-diffusive model. In particular, 
the manifested crossover behavior in the effective mean-square radius 
exponents reveals that slow crossover behavior in general diffusive
multi-species reaction systems may be ascribed to the interference of 
multiple length scales and timescales at early times.
\end{abstract}

\maketitle


\section{Introduction \label{sec:1}}
Nonequilibrium systems exhibiting active-to-absorbing phase transitions
are fundamentally important for understanding a large variety of natural
phenomena
\cite{hinrichsen2000,odor2004,marro2005,odor2008,henkel2008,tauber2014}. 
Among the relevant models, the 
susceptible-infected-recovered (SIR) model for the spread of
epidemic disease in an ensemble of living beings
\cite{kermack1927,*bailey1975,*murray2002,*anderson2013,*pastor2015}, or the 
spread of a non-conserved agent in broader contexts (e.g.~forest fires
\cite{stauffer2018}, chemical reactions \cite{simon2020}, and sociology
\cite{zhao2013}), has long been extensively studied. This 
model and its numerous variants have been applied to the most
varied forms of epidemics
\cite{amaku2003,*hartley2006,*zou2010,*kuhnert2014,*kim2020}, and more
recently have been attracting a surge of attention due to the COVID-19 
pandemic \cite{calafiore2020,*walker2020,*mukhamadiarov2021,odor2021,*odor2021e}. 

The essence of the model assumes that the 
individuals can be categorized into susceptible ($S$), infected ($I$), 
and recovered ($R$) states so that the unidirectional process $S\to I\to R$
occurs, upon the assumption that the infected agent can not pop up
spontaneously but transmits the disease exclusively upon encounter
of $S-I$ pairs ($S+I\xrightarrow{\lambda}2I$), while infected individuals 
recover ($I\xrightarrow{\mu}R$) and cannot revert to a susceptible state 
in any rate. By assuming perfect immunization, the SIR process, which is 
also often referred to as the \textit{general epidemic process} (GEP) 
\cite{grassberger1983}, is deeply connected to the bond percolation process 
both on lattices
\cite{grassberger1983,kuulasmaa1984,janssen1985,cardy1985,janssen2005field} 
and on networks
\cite{moore2000,*newman2002a,*newman2002b,*sander2002,*kenah2007}. Owing to the 
competition of the two sub-processes, the SIR process manifests a continuous
nonequilibrium phase transition that separates the infection dominant
regime, where the epidemic spreads infinitely in the thermodynamic limit, and 
the recovery dominant regime, where the system becomes trapped in an 
\textit{absorbing state} after some time, characterizing the extinction
of the epidemic. 

In the last few decades, voluminous numerical simulation
\cite{grassberger1983,grassberger1997,munoz1999,dammer2004,souza2010,tome2010,souza2010,souza2011}
and field-theoretic \cite{janssen1985,cardy1985,janssen2005field}
analyses have profoundly established the critical properties of this 
transition on $d$-dimensional lattices to be exactly mapped to the 
dynamical isotropic percolation (DIP) 
universality class, except for a subtle difference in their respective local 
cluster growth probabilities \cite{tome2010}. Moreover, the DIP fixed
point remains stable even if partial immunization is implemented, until
the model is tuned into the SIS model which belongs to the directed
percolation (DP) class
\cite{janssen1985,grassberger1997,jimenez2003,dammer2004}, or, according
to the Harris \cite{harris1974} or Harris-Barghathi-Vojta
\cite{barghathi2014,*schrauth2018} criteria, if certain quenched spatial 
disorders or topological disorders are incorporated 
\cite{santos2020,*ferraz2022,*ruslan2022}, as long as the transition is not
destroyed \cite{sander2003} and the dimensionality is not altered
\cite{odor2021}.

As a crucial ingredient for mapping to the DIP, all the previous 
studies tacitly assumed that at least the immune individuals are immobile 
\footnote{Most
previous studies assumed all individuals are immobile, permitting faster
simulation methods. The spread of the epidemic gives rise to an 
``effective diffusion'' for the infected species \cite{janssen2005field}}
which in turn enables 
great simplifications in problem formulation 
\cite{grassberger1983,cardy1985,janssen1985,janssen2005field}. However,
similar to the pair contact process with diffusion (PCPD) \cite{odor2000,*odor2001,*odor2003,*henkel2004}
and the diffusive epidemic process (DEP) \cite{polovnikov2021},
the effects of diffusion shouldn't be overlooked \cite{okubo2001,odor2021}, 
because realistic immune individuals indeed hop around to augment the
mixture of the population, so that such model could find wide applications 
in epidemic spreading among wild grazing and forest animals 
\cite{fofana2017} as well as the spread of human diseases such as whooping 
cough \cite{chinv2010} and 
COVID-19 (when lock-down measures are imposed to cut out most long-range links
and the population is restricted to local mobility), or even in 
autocatalytic reactions with catalyst degradation \cite{simon2020}, to name just 
a few.
Already after taking into account spatial inhomogeneities with the local
densities and the diffusion terms for the SIR rate equation
\begin{IEEEeqnarray}{rCl}
	\partial_t S(x,t) &=& D_{S} \nabla^2 S(x,t) - \lambda S(x,t)
	I(x,t)\,,\nonumber \\
	\partial_t I(x,t) &=& D_{I} \nabla^{2} I(x,t) + \lambda S(x,t) I(x,t)
	-\mu I(x,t)\,,\nonumber \\
	\partial_t R(x,t) &=& D_{R} \nabla^2 R(x,t) + \mu I(x,t)\,,
	\label{eqs:rateeq}
\end{IEEEeqnarray}
it has been shown that the dynamic behavior of such coupled (partial)
differential equations depends on the diffusion rates for both systems with
homogeneous \cite{chinv2010} and inhomogeneous \cite{sakag2021}
couplings. 

What is more, inclusion of diffusion for immune agents may also provoke
nontrivial modifications to the critical properties, as remarked by Janssen 
\textit{et al.} in Ref.~\cite{janssen2004}. First and foremost, at criticality, the 
recovered debris left by starting from a single infectious seed build up a fuzzy 
pattern, in stark contrast to the fully connected percolating cluster in the 
non-diffusive case (see Fig.~\ref{fig:snapshots}). More profoundly, 
from a field-theoretic point of view, after 
casting ``$I$'' and  ``$R$'' into the coarse-grained $\sI(x,t)$
and $\sR(x,t)$ fields \footnote{The local densities in
Eq.~\eqref{eqs:rateeq} are measured for a particular mesoscopic scale, while
the fields in \eqref{eqs:action} can be arbitrarily rescaled in the
continuum limit and can carry anomalous dimensions.} in the continuum 
limit, along with the corresponding response fields $\tilde{\sI}(x,t)$ and 
$\tilde{\sR}(x,t)$, the ensuing bosonic field theory action 
\cite{janssen2005field} (see Appendix~\ref{appA} for the derivation)
\begin{IEEEeqnarray}{rCl}
	\mathcal{A}=\int \ud^dx \ud t \Big\{& &
		\underbrace{\tilde{\sI}\left[\partial_t-D_I\left(\tau-\nabla^2\right) 
		+\frac{g}{2}\left(2\sR-\tilde{\sI}\right)
\right]\sI}_{\mathrm{DIP}} 
	\nonumber \\
	& & + \tilde{\sR} \left(\partial_t- D_R
	\nabla^2\right)\sR -\tilde{\sR} \sI \Big\}\,,
 \label{eqs:action}
\end{IEEEeqnarray}
where $\tau$ denotes the control parameter, renders the duality symmetry 
\begin{equation}
	\tilde{\sI}(x,t) \leftrightarrow -\sR(x,-t)
	\label{eqs:sym}
\end{equation}
no longer held. Note that the very existence of the DIP transition is
induced by the spontaneous breaking of this symmetry \cite{janssen1985},
which arises only if $D_R=0$ \cite{janssen1985,janssen2005field} (see
Sec.~\ref{sec:2}). Once diffusion
for the immune agents sets in, the duality symmetry associated with this 
local accumulation is lost. The full action then
describes a reaction-diffusion type model involving the active species $I$
and the inert species $R$: $I+\emptyset\to 2I$, $I\to R$, in conjunction
with individual diffusion of rates $D_I$ and $D_R$ and reactions for 
particle number restrictions in a bosonic representation; see
Eq.~\eqref{eqs:allreact}.

While the rate equation system \eqref{eqs:rateeq} yields qualitatively good
predictions for the evolving behavior of an epidemic process, it still amounts 
to a mean-field treatment in which correlations in the infection interactions had 
been factorized. Hence, it ignores spatiotemporal fluctuations and
correlations of the reaction processes that increasingly become
crucial for low-dimensional
systems near criticality \cite{hinrichsen2000}. To fully account for
the effects of fluctuations and correlations when individual diffusion
is also present in the SIR model, one can resort either to a
field-theoretic analysis \cite{tauber2014,janssen1985,janssen2005field}, which 
is usually quite challenging for more complicated models, or to a
straightforward implementation of the stochastic reactions via simulations
\cite{grassberger1983,tome2010}. In this work, we will take the latter approach 
to study the critical properties of the diffusive SIR (DSIR) process on a 
two-dimensional lattice. Field theory action is primarily introduced to give a 
more profound motivation for this study. Yet, we hope the numerical results 
could be beneficial to further advancing field-theoretic analyses as well.

The reminder of this paper is organized as follows: In the next section,
we give a detailed exposition on the violation of the duality symmetry
in the DSIR. We then detail our simulation method in Sec.~\ref{sec:3}
and compute various critical exponents both in the dynamical regime and
in the stationary state in Sec.~\ref{sec:4}. Finally, Sec.~\ref{sec:5} 
summarizes this work and provides a brief outlook.

\section{Violation of the duality symmetry \label{sec:2}}
The response field $\tR(x,t)$ only appears linearly in the action
\eqref{eqs:action}. Hence, it can be integrated out from the path
integral to retain only the DIP part of action \eqref{eqs:action}, 
which is equivalent to computing the functional derivative
$\frac{\delta \mathcal{A}}{\delta\tR}=0$ \cite{tauber2014} that leads
to a constraint for the $\sR(x,t)$ field
\begin{equation}
	\partial_t \sR=D_R \nabla^2 \sR + \sI\,.
	\label{eqs:rcons}
\end{equation}
Without the precence of diffusion for the immune agents, we simply have
\begin{equation}
	\sI(x,t)=\partial_t \sR(x,t) \,\,\, \text{and} \,\,\,
\sR(x,t)=\int_{-\infty}^{t}\!\ud t'\sI(x,t')\,,
\label{eqs:DIPRI}
\end{equation}
with which the DIP part of the action \eqref{eqs:action} can be further
manipulated through integrating by parts and becomes \cite{janssen1985}
\begin{IEEEeqnarray}{rCl}
\mathcal{A}_{\mathrm{DIP}}\!&=&\! \int\!\ud^dx\ud t \tI\left[\partial_t-D_I
	\left(\tau - \nabla^2\right) + \frac{g}{2}\left(2\sR -
\tilde{\sI} \right) \right]\partial_t \sR \nonumber \\
&=&\int\!\ud^d x \ud t\Big\{\partial_t\sR[-\partial_t + D_{I}(\tau -
\nabla^2)] \tI \nonumber \\ 
&& \qquad \qquad - \frac{g}{2}\sR^2 \partial_t \tI -\frac{g}{2}\tI^2 \partial_t
\sR\Big\}\,. 
\label{eqs:DIPtrans}
\end{IEEEeqnarray}

Now we apply the duality transformation Eq.~\eqref{eqs:sym}
[$\tilde{\sI}(x,t) \leftrightarrow -\sR(x,-t)=-\int_{-\infty}^{-t}\!\ud
t' \sI(x,t')$] on $\mathcal{A}_{\mathrm{DIP}}$ to substantiate that the
DIP action is invariant after transformation:
\begin{IEEEeqnarray}{rCl}
	\mathcal{A}_{\mathrm{DIP}}'&=& \int\!\ud^d x\ud t\Big\{ \left(
			\partial_t[-\tI(-t)]\right)[- \partial_t -
			D_I(\tau-\nabla^2)]\nonumber \\
			&&\qquad  \times [-\sR(-t)] -
			\frac{g}{2}[-\tI(-t)]^2\left(
			\partial_t[-\sR(-t)] \right)\nonumber \\
			&& \qquad -\frac{g}{2}[-\sR(-t)]^2\left(
		\partial_t[-\tI(-t)] \right)\Big\}\nonumber \\
	&\stackrel{t'=-t}{=}& \int\!\ud^d x\ud t'\Big\{
	\partial_{t'}\sR(t')[-\partial_{t'}+D_{I}(\tau-\nabla^2)] \tI(t')
	\nonumber\\
	&&\qquad \quad - \frac{g}{2}\tI(t')^2\partial_{t'}\sR(t') -
\frac{g}{2} \sR(t')^2\partial_{t'}\tI(t')\Big\}\nonumber\\
&\stackrel{\text{Eq.~\eqref{eqs:DIPtrans}}}{=}& \mathcal{A}_{\mathrm{DIP}}\,,
\label{eqs:symeq}
\end{IEEEeqnarray}
where in the second equation we have again employed integration by parts.

The above derivation demonstrates that, after integrating out the
response field $\tR$, it is crucial for Eq.~\eqref{eqs:DIPRI} to be strictly 
valid for the DIP part of action \eqref{eqs:action} to be symmetric under the 
duality transformation. According to Eq.~\eqref{eqs:rcons}, implementing 
diffusion for the immune agents immediately spoils this requirement.

\section{Simulation method \label{sec:3}}
To verify the effects of diffusion on the critical properties,
in this paper, we perform large-scale Monte Carlo simulations 
for the simplest $D_R=D_I=D_S=D$ case on a 
two-dimensional square lattice, with each site being occupied by 
exactly one individual. Since individuals constantly jiggle around after 
the incorporation of 
diffusion, efficient simulation methods that had been exploited for the
non-diffusive SIR (hereafter referred to as SIR), such as 
traversing over a linked list for active agents \cite{dammer2004}, are 
not feasible in a
straightforward way. Henceforth, to reduce the simulation overhead, we
resort to the sequential updating scheme \cite{hinrichsen2000,odor2021}. 
At each Monte Carlo (MC) time step $t$, it is preferable to implement 
reactions and individual diffusions into two separate sequential sweeps 
to prevent diffusion events from interfering with reaction events. The 
system is then updated as follows.

(1) Two arrays $\mathbf{x}$ and $\mathbf{X}$ (initially, $\mathbf{x}=
\mathbf{X}$) of size $N=L\times L$, representing the current
and the future states of the system, are maintained to trace the
updates. At each site, the state variables $x_i$ and $X_i$ can
assume one of the $S$, $I$, and $R$ states. During the
reaction sweep, depending on the state value $X_i$ of a selected site 
$i$, the state is updated either from $X_i=S$ to $X_i=I$ with 
probability $\lambda$ if the selected nearest neighbor $j$ for 
contact is in the state $x_j=I$, or from $X_i=I$ to $X_i=R$ with 
probability $\mu=1-\lambda$; otherwise, $X_i$ remains intact.

(2) In the diffusion sweep, we simply swap the states of a selected site
$i$ and its randomly selected neighbor $j$ with respect to identical 
probability $D$: $X_i\xleftrightarrow{D}X_j$. One caveat to note is that
the conventional rightward-downward sweep tends to cause a biased diffusion 
towards the right and the down directions. To counteract this artifact 
so that individuals diffuse unbiasedly, the lattice is swept 
alternately in a forward manner for odd time steps, and in a backward 
manner for even time steps.

(3) Set $\mathbf{x}=\mathbf{X}$ after each cycle of the above two sweeps
and increase $t$ by one to start a new updating cycle, until the
prescribed/conditioned simulation time is reached.

In principle, the artifact in step 2 can be circumvented with a greater
effort by matching the pairs for swapping through a domino tiling 
\cite{kenyon2009}, permitting a parallel update of the system. Nevertheless, 
the entirety of the above updates should be considered as happening 
simultaneously for each time step. As a side remark, note that
critical properties of a system are governed by the emerged long-range 
correlations and are insensitive to microscopic details, as exemplified
by the bosonic representation of the DSIR in Appendix~\ref{appA}, in
which a site can even contain more than one individual. In this respect,
any mechanism that provides isotropic \textit{local} mobility of the 
individuals can be defined as a valid diffusion action, and the above 
implementation of diffusion for our fully occupied lattice via simple 
state swapping is justified.

\begin{figure}[!tbp]
	\centering
	\includegraphics[width=0.43\textwidth]{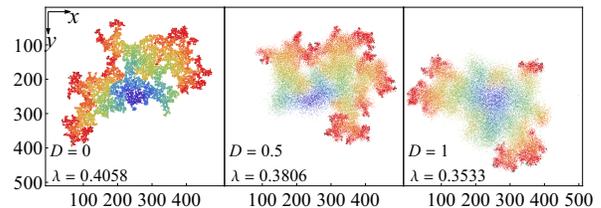}
	\caption{Snapshots for the critical SIR process with diffusion 
	rates $D=0, 0.5,$ and $1$ on a $500 \times 500$ lattice. The
	$S$ and $I$ species are colored in white and black. The
	rainbow spectrum beared by the $R$ species, from blue to red, 
	linearly marks their relative generating time.}
	\label{fig:snapshots}
\end{figure}

In order to simulate the critical dynamics of absorbing phase transitions, 
there are two conventional ways of initializing the runs,
i.e.~homogeneously filling the lattice with the active agents or
starting each run from a single active seed \cite{hinrichsen2000}. 
Due to the irreversible nature of the dynamics, the former only
amounts to the relaxational process for recoveries. Therefore, it is 
customary to study the spreading dynamics by initializing the lattice 
with a single infectious seed \cite{hinrichsen2000,henkel2008,odor2021,
grassberger1979} 
placed at the center of the lattice; cf.~the typical clusters
obtained at criticality shown in Fig.~\ref{fig:snapshots}. The growth of
clusters is characterized by the number of $I$ individuals 
$N_I(\lambda,t)$; the survival probability
$P_{\mathrm{sur}}(\lambda,t)$ \footnote{A run up to time $t$ is 
surviving provided $N_I(t)> 0$.}, with both quantities averaged over all
runs; and the mean-square radius $R_I^2(\lambda,t)$, averaged over
survival runs. Simulations were terminated if the distance of any
$I$ or $R$ individual away from the center exceeds $L/2$. Above the
transition point $\lambda_c$, after denoting $\Delta=\lambda-\lambda_c$,
the following scaling relations hold
\cite{henkel2008,janssen2005field}
\begin{IEEEeqnarray}{rCl}
	N_I(\lambda,t)&=&t^{\theta_I}\hat{N}_I\left(
	\Delta^{\nu_{\parallel}} t \right)\,, \IEEEyesnumber*
	\IEEEyessubnumber* \label{eqs:NI} \\
	P_{\mathrm{sur}}(\lambda,t)&=&t^{-\delta}\hat{P}_{\mathrm{sur}}\left(
	\Delta^{\nu_{\parallel}}t \right)\,, \\
        R_I^2(\lambda,t)&=&t^{Z_I}\hat{R}_I^2\left(
	\Delta^{\nu_{\parallel}}t \right)\,,
\end{IEEEeqnarray}
giving rise to the power laws at criticality,
\begin{equation}
	N_I(t)\sim t^{\theta_I}, \quad P_{\mathrm{sur}}(t)\sim
	t^{-\delta}, \quad R_I^2(t)\sim t^{Z_I},
	\label{eqs:power}
\end{equation}
where $\theta_I$, $\delta$, and
$Z_I=2/z_I=2\nu/\nu_{\parallel}$ are spreading exponents, whilst
$\nu$ and $\nu_{\parallel}$ are related to the correlation length
and the characteristic time, diverging as $\xi\sim \Delta^{-\nu}$ 
and $t_c\sim \xi^{z_I}\sim \Delta^{-\nu_{\parallel}}$. For 
two-species systems, it is also appropriate to define the
counterparts $N_R(t)\sim t^{\theta_R}$ and $R_R^2(t)\sim t^{Z_R}$, for
the $R$ species \cite{deng2020}.

\section{Critical exponents \label{sec:4}}
In this section, we show that after a crossover the dynamical 
spreading exponents of the DSIR consistently deviate from the SIR/DIP, 
resulting in an altered hyperscaling relation. In addition, finite-size 
scaling analyses of the stationary state further corroborate the main 
conjecture.

\begin{figure}[!tbp]
	\centering
	\includegraphics[width=0.485\textwidth]{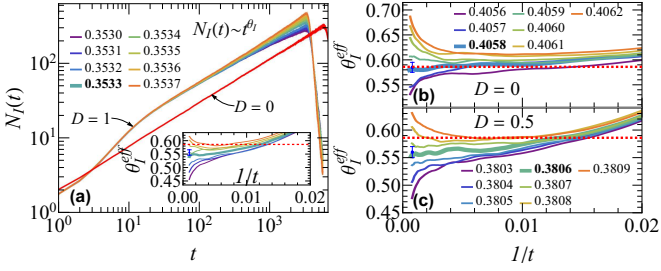}
	\caption{Growth of the $I$ population $N_I(t)$ from a single
		infectious seed on a $L=4001$ square lattice in the
		vicinity of criticality for (a) $D=1$, and the
		evolution of the corresponding effective exponent
		$\theta_I^{\mathrm{eff}}$ in the inset, in (b) for 
		$D=0$, and in (c) for $D=0.5$. For comparison, the 
		red solid line in (a) depicts the critical $N_I(t)$ 
		result for $D=0$. The critical points, emphasized by the
		thick curves, are estimated to
		$\lambda_c\simeq 0.4058(1)$, $\lambda_c\simeq
		0.3806(1)$, and $\lambda_c\simeq 0.3533(1)$ 
		for $D=0$, $D=0.5$, and $D=1$, respectively. The horizontal 
		dashed red line indicates
	the DIP value; cf.~Table~\ref{tab:exps}. All results were averaged 
	over $10^4$ independent runs.}
	\label{fig:NI}
\end{figure}

As shown in Fig.~\ref{fig:NI}(a), in stark contrast to the SIR, in
which a pure power law is manifested for $N_I(t)$ at
$\lambda_c$, the DSIR process exhibits a crossover before an
asymptotic scaling regime is approached. Owing to the competition of
diffusive spreading and local reactivity, this crossover can be separated
into two stages: first, the initially abundant $S$ content renders the
kinetic to be \textit{reaction-limited} and once the established correlations 
exceed the lattice spacing at $t \sim 10$, the enhanced mixture of $S$ and 
$I$ populations kicks in a boosted spread; then at large times, the produced 
$R$ debris effectively dilute the local reactant densities and the system 
becomes \textit{diffusion-limited} \footnote{The dilution of local
reactant densities by $R$ individuals would be noticeable
only at the critical point though, as for $\lambda > \lambda_c$
one would have a circular infectious front, which will be always ahead 
of $R$ individuals.}.
In finite systems, this process goes on until reachable $S$ individuals 
are depleted. The observed asymptotic scaling behavior permits one to 
estimate the transition point $\lambda_c$ by observing the evolution of 
the local slope, i.e.~the effective exponent
\begin{equation}
\theta_I^{\mathrm{eff}}=\left\lvert\frac{\ln\left[ N_I(t)/N_I(t/b)] \right]}{\ln(b)}
\right\rvert,
	\label{eqs:eff}
\end{equation}
with $b>1$ \footnote{We used $b=1.4$ and employed moving average to smooth 
out local fluctuations}; similarly, $\theta_R^{\mathrm{eff}}$, $\delta^{\mathrm{eff}}$, 
$Z_I^{\mathrm{eff}}$, and $Z_R^{\mathrm{eff}}$ can be defined. 
The transition point is then identified by spotting
the $\lambda$ value that gives rise to an asymptotically stationary
$\theta_I^{\mathrm{eff}}$. Fig.~\ref{fig:NI} illustrates that on the one hand, 
the transition point decreases with increasing diffusion rates; on the
other hand, while $\theta_I\approx 0.585(10)$ for $D=0$ conceivably
recovers the DIP value, the DSIR exponent 
value $\theta_I\approx 0.55(2)$ and $0.56(1)$ for
$D=0.5$ and $D=1$ demonstrate a consistent, albeit slight, deviation 
from the DIP. This downward shift in $\theta_I$ can be understood
as a consequence of the above-mentioned dilution effect which mitigates 
the infections at the infectious front to some extent. 

\begin{figure}[!tbp]
	\centering
	\includegraphics[width=0.485\textwidth]{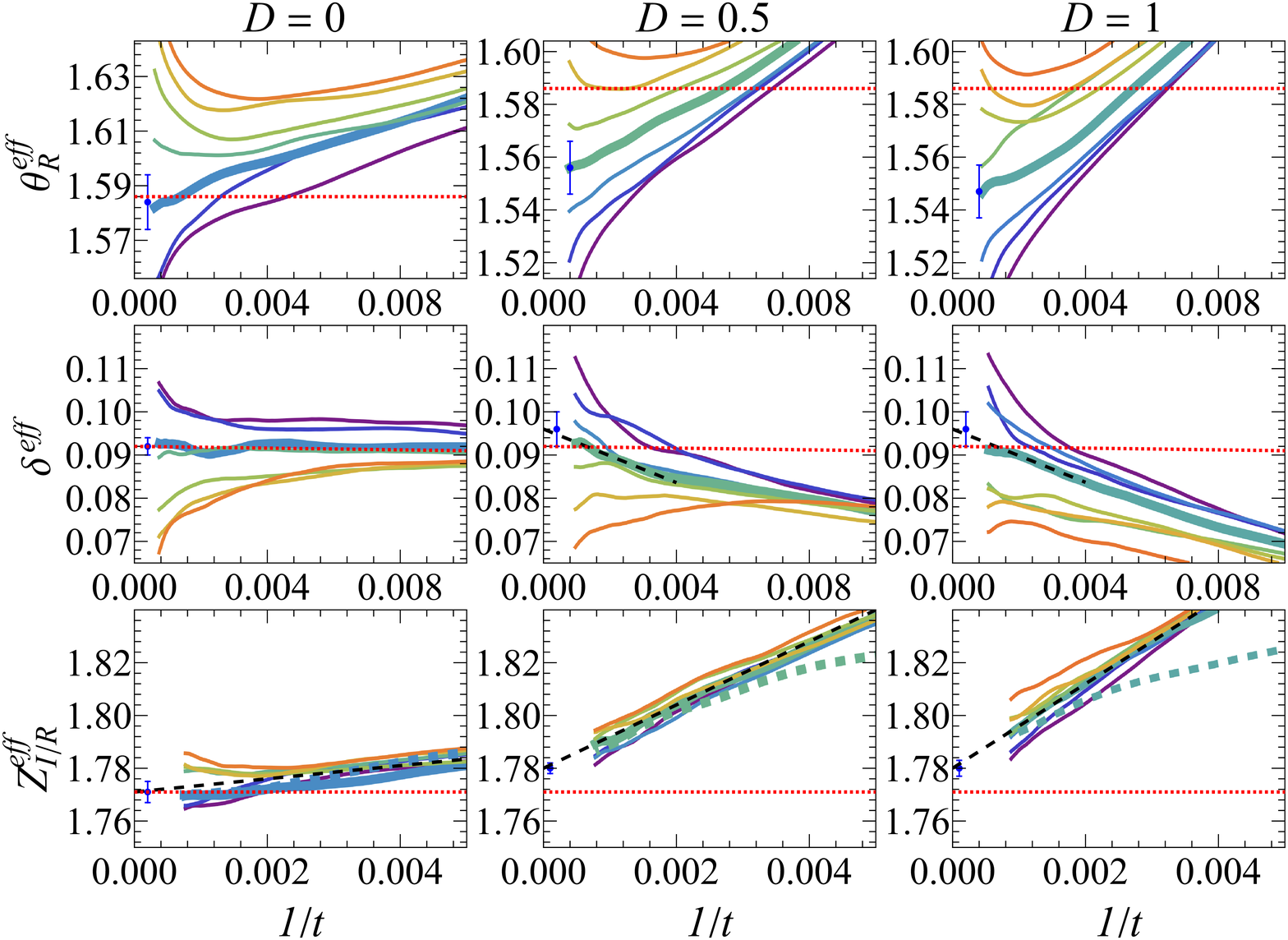}
	\caption{The effective exponents $\theta_R^{\mathrm{eff}}$ (top), 
	$\delta^{\mathrm{eff}}$ (middle), as well as $Z_I^{\mathrm{eff}}$ and 
$Z_R^{\mathrm{eff}}$ (bottom) vs.~$1/t$ for $D=0,0.5$, and $1$. The
solid thick lines correspond to the respective critical points. In the
bottom panes, the dashed thick curves represent the critical
$Z_R^{\mathrm{eff}}$, while the dashed black lines extroplate the
effective exponents to $1/t\to 0$. The standard DIP values are marked by the
horizontal dashed red lines; cf.~Table~\ref{tab:exps}.}
	\label{fig:eff}
\end{figure}

Figure \ref{fig:eff} further unambiguously shows departures of other
DSIR spreading  exponents from those of the SIR, which again align
with the DIP. In particular, on top of
$\theta_I^{\mathrm{eff}}$, which already manifests an evident crossover 
behavior in the DSIR (cf.~Fig.~\ref{fig:NI}), all the remaining 
effective exponents for DSIR are displaying even more 
pronounced crossover behaviors as compared to the SIR. 
Hence, $N_I(t)$ is a more apt observable for critical point estimation, 
even though it is still quite nontrivial, similarly to the 
PCPD, to fully take into account the corrections to
scalings~\cite{odor2000,*odor2001,*odor2003,*henkel2004,park2014}, 
thereby the exponents also seem 
to exhibit a slight dependence on the implemented diffusion rate 
\footnote{Although the DSIR exponents only slightly deviate from
the DIP values, note that the most precise decay exponent of
the PCPD also differs from that of the pair contact process
(PCP, i.e.~DP) by just $\sim 0.014$ \cite{park2014}. Since the $R$ species is
inert, we speculate that the DSIR is less affected by corrections to 
scaling as compared to the PCPD, as suggested by the asymptotically
identical values of $Z_{I}^{\mathrm{eff}}$ and $Z_{R}^{\mathrm{eff}}$;
also cf.~Fig.~5 of Ref.~\cite{deng2020}. Ref.~\cite{ruslan2022}
studied a similar DSIR process with a critical population
density and with a relatively smaller system size, it seems these slight 
deviations were not captured due to their larger error margins.};
cf.~Table~\ref{tab:exps}.
Yet an important observation to make is that the effective spreading 
exponents $Z_I^{\mathrm{eff}}$ and $Z_R^{\mathrm{eff}}$, while closely 
clinging to each other since early times for $D=0$, only close up their 
noticeable gap and converge to an identical value $Z_I=Z_R=Z$
asymptotically for the DSIR, suggesting that there is really 
just one unique set of length scale $\xi$ and timescale $t_c$, that 
renders a critical system scale invariant. Since $Z_I^{\mathrm{eff}}$ and
$Z_R^{\mathrm{eff}}$ are related to the effective correlation exponents,
the manifested crossover behavior at early times can then be ascribed 
to the interference of multiple length and time scales, resulting from 
diffusion ($\ell_D\sim t^{1/z_D}=t^{1/2}$) and the cutoffs
of the correlation functions $\angs{\sI(r,t)\tilde{\sI}(0,0)}$ and 
$\angs{\sR(r,t)\tilde{\sI}(0,0)}$, until the dominant scales $\xi$ and 
$t_c$ are singled out as $t\to \infty$, masking processes with shorter 
characteristic length scales and timescales. Note that $z=2/Z<z_D=2$,
so the system is superdiffusive and $\xi$ is bound to be larger than 
$\ell_D$.

The DIP spreading exponents are related by the hyperscaling relation
\cite{munoz1997} 
\begin{equation}
	\theta_I=\frac{d Z}{2}-2\delta -1\,.
	\label{eqs:hyp}
\end{equation}
Furthermore, the size of the immune cluster should grow linearly 
with the linear extension of the cluster as $\xi^{d_{f}}\sim t^{d_f Z/2}$
in a surviving run, where $d_f$ denotes the fractal dimension
\cite{stauffer2018}. By utilizing the hyperscaling relation
$d_f=d-\beta/\nu$ and the scaling relation 
$\delta=\beta/\nu_{\parallel}=\beta Z/2\nu$ \cite{stauffer2018,janssen2005field} 
(see below for the definition of $\beta$), the average size of the immune 
cluster for all runs is then obtained by further multiplying the 
expression with the survival probability
\begin{equation}
	N_R(t)\sim t^{dZ/2 - 2\delta}\,.
	\label{eqs:DIPR}
\end{equation}
Eqs.~\eqref{eqs:hyp} and \eqref{eqs:DIPR} suggest that 
$\theta_R=\theta_I+1 \simeq 1.586$ for the DIP. The SIR is 
naturally in full compliance with these relations. 

\begin{table}[!t]
\scriptsize
\centering
	\setlength{\tabcolsep}{0.125em}
	\caption{Critical (and scaling) exponents of the DIP \cite{grassberger1983,munoz1999,souza2011}, 
	the SIR ($D=0$), and the DSIR ($D=0.5,\,1$)
\protect\footnote{Uncertainties in the last digit were estimated from the
effective exponents for the spreading exponents and from fitting errors
for the remaining ones.}.}
	\begin{tabularx}{0.485\textwidth}{l
			  S[table-format=1.6]
		          S[table-format=1.6]
		          S[table-format=1.6]
		          S[table-format=1.6] 
		          S[table-format=1.5]
		          S[table-format=1.6]
		          S[table-format=1.6]}
		\toprule
		$\,\,\,D$ & $\theta_I$ & $\theta_R$ & $\delta$ &
		$Z$ & $\nu_{\parallel}$ & $\beta/\nu$ &
		$\gamma/\nu$  \\
                \midrule
		DIP & 0.586 & 1.586 & 0.092 &
		1.771 & 1.5057 & 0.1042 & 1.792 \\
		$0$ & 0.585(10) & 1.584(10) &
		0.092(2) & 1.771(4) & 1.51(1) &
		0.1040(2) & 1.810(2) \\
		$0.5$ & 0.56(1) & 1.55(1) &
		0.096(4) & 1.780(2) & 1.46(1) &
		0.096(2) & 1.764(4) \\
		$1$ & 0.55(2) & 1.54(1) &
		0.096(4) & 1.780(3) & 1.47(1) &
		0.093(3) & 1.755(3)\\
		\bottomrule
        \end{tabularx}
 \label{tab:exps}
\end{table}

Now in the DSIR, although the relation $\theta_R\simeq \theta_I+1$
still seems to be valid, Eq.~\eqref{eqs:DIPR} predicts $\theta_R\approx
1.588(9)$, which is higher than the obtained values
$1.55(1)$ and $1.54(1)$, implying that the scaling
relations $d_f=d-\beta/\nu$ and/or $\delta=\beta/\nu_{\parallel} 
\beta Z/2\nu$ utilized for Eq.~\eqref{eqs:DIPR} may not be held for the
DSIR. Nonetheless, by taking into account the dilution effect, the 
density deduced from Eq.~\eqref{eqs:DIPR} should be further reduced by a 
factor of $t^{-(\theta_{I.\mathrm{DIP}}-\theta_{I.\mathrm{DSIR}})}$,
leading
\begin{equation}
	\theta_R\simeq
	\frac{dZ}{2}-2\delta-(\theta_{I.\mathrm{DIP}}-\theta_{I.\mathrm{DSIR}})
	\approx 1.55(2)
	\label{eqs:thetaR}
\end{equation}
to be compatible with numerical values within error margins.
In addition, the absence of the duality symmetry leads to
an apparent violation of the hyperscaling relation Eq.~\eqref{eqs:hyp} in 
the DSIR. To break down this discrepancy, we need to take into
consideration the renormalization corrections to the naive scaling
dimensions, whereupon, when expressed in terms of an arbitrary time
scale $T\sim \kappa^{-2/Z}$, we have $[\sI]\sim T^{-dZ/4-1/2+\rho}$ and
$[\tilde{\sI}]\sim T^{-dZ/4+1/2+\chi}$, so that $N_I(t)\sim
\angs{\int d^dx\sI(x,t)\tilde{\sI}(0,0)}\sim t^{\rho+\chi}$ holds,
where $\rho$ and $\chi$ are the anomalous dimension of the fields.
Furthermore, in field theories with absorbing states, one has
$[\tilde{\sI}]\sim T^{-\delta}$ \cite{munoz1997}. Hence, given the
symmetry $\tilde{\sI}(x,t) \leftrightarrow - \int_{-\infty}^{-t} dt' 
\sI(x,t')$ which renders $\rho=\chi$, Eq.~\eqref{eqs:hyp} immediately 
follows in the DIP, whereas in the DSIR, only $\chi$ can be eliminated
and the hyperscaling relation is altered to
\begin{equation}
	\theta_I=\frac{dZ}{4}-\delta-\frac{1}{2}+\rho\,.
	\label{eqs:hypd}
\end{equation}
Inserting other exponent values, the DSIR value $\rho\approx
0.26(2)$ differs from the DIP value $\rho=\theta_I/2=0.293$.

\begin{figure}[!tbp]
	\centering
	\includegraphics[width=0.46\textwidth]{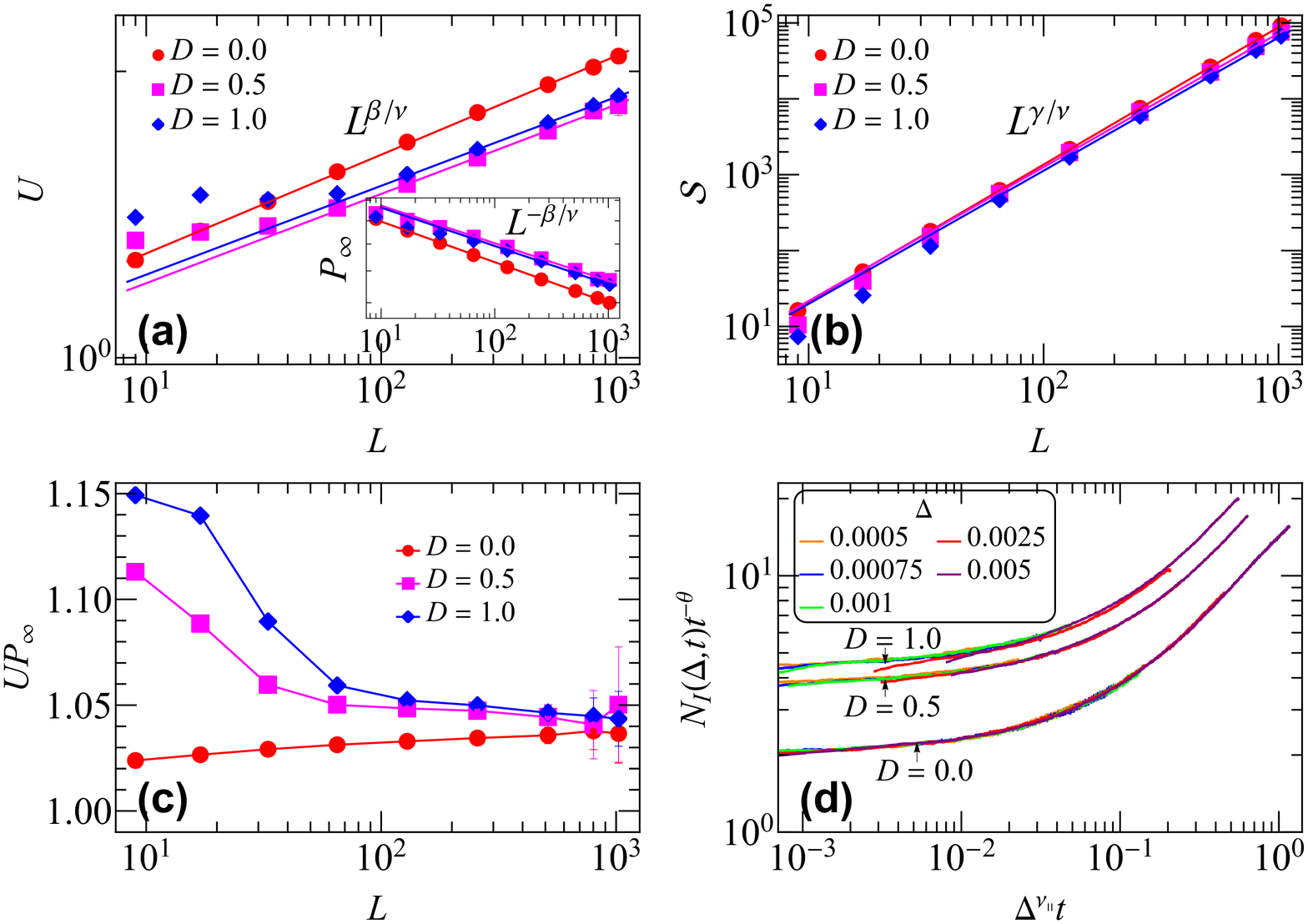}
	\caption{The finite-size scaling results of (a) $U$ and
		$P_{\infty}$ (inset), (b) $\mathcal{S}$, and (c) $UP_{\infty}$; 
		and (d) the data collapse results for $N_I(\Delta,t)$.
		The results in (a)--(c) were averaged from $10^4$ to 
		$10^8$ runs, and the results in (d) were obtained with $L=2001$, 
	        averaged over $10^4$ runs.}
	\label{fig4}
\end{figure}

Starting from a seed, the peculiar DSIR dynamical spreading behavior at 
criticality eventually results in many shattered remnant clusters 
(cf.~Fig.~\ref{fig:snapshots}), rather than a fully connected percolating 
cluster as in the DIP/SIR 
\cite{grassberger1983,tome2010}. Consequently, the DIP exponents 
$\beta$ and $\gamma$, associated with the percolation probability 
$P_{\infty}\sim \lvert\Delta\rvert^{\beta}$ and the mean cluster size 
$\mathcal{S}\sim \lvert\Delta\rvert^{-\gamma}$, respectively, are not well defined
in the DSIR. Nevertheless, in the sense 
of how one infected seed may affect a sizable population, we can define 
the ``mean cluster size'' as the average eventual number of recovered 
individuals $\mathcal{S}=\angs{N_{R\infty}}$ after the disease dies out, 
then the definitions for the corresponding second moment 
$\mathcal{M}=\angs{N_{R\infty}^{2}}$ 
and the cumulant $U=\mathcal{M}/\mathcal{S}^2$ follow subsequently. 
Furthermore, for systems 
with a definite size, the DSIR ``percolation probability'' $P_{\infty}$ 
can as well be understood as the fraction of 
runs with any $I$ individuals ever reached the border. Similar to the 
SIR, these observables are expected to follow the following finite-size
scalings at criticality \cite{de2011new}
\begin{equation}
	\mathcal{S}\sim L^{\gamma/\nu}\,, \quad U\sim
	L^{\beta/\nu}\,, \quad P_{\infty}\sim
	L^{-\beta/\nu}\,,
	\label{eqs:FSS}
\end{equation}
suggesting $UP_{\infty}=\mathrm{const.}$

Figures~\ref{fig4} (a)--(b) do justify the above finite-size scalings
for large system sizes. However, these scalings are strongly disturbed
by diffusion for smaller system sizes, as evidently shown by the crossover
of $UP_{\infty}$ in Fig.~\ref{fig4} (c). To bridge the dynamical exponents with
the stationary scaling exponents, by utilizing Eq.~\eqref{eqs:NI}, we also 
estimated the exponent $\nu_{\parallel}$ by collapsing the
$N_I(\Delta,t)t^{-\theta_{I}}$ data, for several $\Delta$s, to the scaling 
function 
$\hat{N}_I(\Delta^{\nu_{\parallel}}t)$ with respect to the rescaled time
$\Delta^{\nu_{\parallel}}t$. In Fig.~\ref{fig4} (d), by fitting all the 
datasets for different $\Delta$s with an 8th-order polynomial, the best 
estimations for $\nu_{\parallel}$ were obtained by minimizing the sum 
squared error.
Collecting the obtained exponent values in Table~\ref{tab:exps}, we see
that the DSIR exponents, as well as the deduced exponents
$\nu=Z\nu_{\parallel}/2\approx 1.30(1)$, $\beta\approx 1.21(1)$, and 
$\gamma \approx 2.29(2)$, again show deviations from those of
the DIP/SIR: $\nu\simeq 4/3$, $\beta\simeq 5/36$, and $\gamma \simeq 43/18$
\cite{grassberger1983}. What is more, since $\beta$ and $\gamma$ are not 
defined on
connected clusters, the scaling relations 
\cite{grassberger1983,janssen2005field}
$\delta=\beta/\nu_{\parallel}=\beta/\nu\times Z/2$ and
$(2\beta+\gamma)/\nu d=1$ also seem to be violated in the DSIR.

\section{Summary \label{sec:5}}
Our simulations and scaling analyses show that the inclusion
of diffusion for immune individuals profoundly alters the critical
properties of the SIR/DIP in two dimensions. Distinct anomalous 
scaling dimensions emerge due to the absence of the 
duality symmetry, leading to an altered hyperscaling relation. 
In particular, the effective exponents $Z_{I}^{\mathrm{eff}}$ and
$Z_{R}^{\mathrm{eff}}$ indicate signatures of multiple length scales and 
timescales at early times, which qualitatively explain the
manifested crossover behavior. Hence, in addition 
to the PCPD, which may be considered a diffusive coupled two-species system 
\cite{deng2020} and which also demonstrates a slight exponent change
\cite{Note6}, 
the DSIR provides another example of how diffusion 
may introduce a singular perturbation, characterized by a slow crossover 
behavior, to an otherwise well-behaved 
multi-species system and as in the PCPD \cite{odor2000,*odor2001,*odor2003,*henkel2004,deng2020}. Such 
perturbation may lead to even more intricate dynamics if there are more 
than one active species. Except for some multi-species directed 
percolation processes \cite{tauber2014,janssen1997,*tauber1998,*tauber2012}, 
the critical properties of 
general active-to-absorbing transitions that involve higher-order or
multi-species reactions are still scarcely studied and are as yet 
incompletely understood. We hope our work will shed some 
light on these fields. 

For more realistic epidemiologic modeling, such as in metapopulation
models built from internally strongly connected modules
\cite{colizz2008}, structural disorders are strong enough to affect the
critical properties. It is then interesting to investigate how 
critical properties will be affected by diffusion in conjunction with
structural disorders, by constructing long-range links 
\cite{odor2021,ruslan2021r}.

\begin{acknowledgements}
We are very much indebted to R\'{o}bert Juh\'{a}sz, Ruslan Mukhamadiarov,
and Uwe T\"{a}uber for helpful discussions. Support from the Hungarian 
National Research, Development and Innovation Office NKFIH (K128989) is 
acknowledged. Most of the numerical work was done on KIFU supercomputers 
of Hungary.
\end{acknowledgements}

\appendix

\section{Field theory and the action \label{appA}}

In this Appendix, we show how Eq.~\eqref{eqs:action} can be obtained by 
mapping the classical master equation of reaction-diffusion processes 
onto a field theory action via the Doi-Peliti formalism
\cite{doi1976second,doi1976stochastic,peliti1985path} (also see
Refs.~\cite{tauber2005app,cardy2008non,tauber2014} for
more recent reviews). 
To begin with, let us note that the action
Eq.~\eqref{eqs:action} serves to describe the system near the
transition, when the $I$ species is close to extinction so that
the density of the $I$ species is vanishingly small as compared to 
that of the $S$ species. Since $S$ individuals are basically 
everywhere, it suffices to consider them as a background.
Then, similar to the decoupling of predators from preys near 
the predator extinction transition in the Lotka–Volterra model for 
predator-prey systems \cite{tauber2012}, the SIR reactions can be 
replaced with $I\to 2I$ and $I\to R$ by ignoring the existence of 
the $S$ species. Alternatively and more straightforwardly, for the
conventional full-lattice setup \cite{grassberger1983,tome2010}, 
in which every lattice site is occupied exactly by one individual of
$S$, $I$ or $R$ state, the vast existing $S$ state can be treated as
the ``vaccum'' state $\emptyset$ as in the contact process for the
lattice susceptible-infected-susceptible (SIS) model
\cite{harris1974con,henkel2008}. 

However, as will become clear later, the Doi-Peliti formalism considers 
particles as ``bosonic'', meaning arbitrarily many particles of either
species could occupy a lattice point. Therefore, in field theory, to 
prevent local particle numbers from diverging in the active phase, one 
can either mimic the mutual exclusion of particles in simulations by 
imposing a hard-core constraint \cite{van2001} or more heuristically 
just add the reaction $I+I\to I$ and, without loss of generality, the 
reaction $I\to\emptyset$, to restrict the local particle numbers 
\cite{tauber2012}. We should remark that retaining either or both of 
these two reactions will lead to the same effective field theory. 
Furthermore, the reaction $I+R\to R$ is added to suppress further 
productions of $R$s from $I$s if an $R$ individual is already present 
at a location. We then consider the following set of reactions in 
the bosonic field theory \cite{janssen2005field}:
\begin{equation}
	I\mathrel{\mathop{\rightleftarrows}^{\lambda}_{\kappa}} 2I\,,\quad
	I\xrightarrow{\sigma}\emptyset\,, \quad I\xrightarrow{\mu}R\,, \quad
	I+R\xrightarrow{\nu}R\,.
	\label{eqs:allreact}
\end{equation}

Henceforth we mainly follow the derivations in Sec.~2.2 of 
Ref.~\cite{janssen2005field}, filling necessary gaps. Suppose 
there is no site occupation number 
restriction, i.e.~we are considering a ``bosonic'' system with a configuration
$\{n,m\}=(\dots,n_i,\dots; \dots, m_i, \dots)$ with $n_i$ particles of 
species $I$ and $m_i$ particles of species $R$~on site $i$, etc., where 
$n_{i},m_{i}=0,1,2\ldots\,$. The integer occupation number changes of each 
species ($I$, $R$) can be accounted for by using the creation and 
annihilation operators $\{\ha, \hb\}$ and $\{a,b\}$ that satisfy the 
\textit{bosonic ladder operator algebra}: 
$[a_{i},a_{j}]=[\ha_{i}, \ha_{j}] = [b_{i},b_{j}]=[\hb_{i}, \hb_{j}]=0\,, \,
[b_{i},\hb_{j}]=[a_{i}, \ha_{j}]=\delta_{ij}$. Denoting $|n_i\rangle$ 
the particle number eigenstate on site $i$ and defining the vacuum state
through $a_{i}|0\rangle=0$, the bosonic algebra dictates that
$a_{i}\left|n_{i}\right\rangle=n_{i}\left|n_{i}-1\right\rangle,\,
a_{i}^{\dagger}\left|n_{i}\right\rangle=\left|n_{i}+1\right\rangle$, and 
$a_{i}^{\dagger} a_{i}\left|n_{i}\right\rangle=n_{i}\left|n_{i}\right\rangle$.
The full state describing a given configuration of the system can then be 
constructed from the vacuum state as the \textit{Fock product state}
$\left|\left\{n,m\right\}\right\rangle =
\prod_{i}\ha_{i}^{n_{i}}\hb_{i}^{m_{i}}|0\rangle$ and the state of the 
entire stochastic system $|\Phi(t)\rangle$ is expressed as a superposition 
of all possible configuration states
$|\Phi(t)\rangle=\sum_{\left\{n,m\right\}}
P\left(\left\{n,m\right\};t\right)\left|\left\{n,m\right\}
\right\rangle$, weighted with the time-dependent configuration probability. 
The master equation for the configuration probability 
$P\left(\left\{n,m\right\};t\right)$ is then cast into an ``imaginary-time 
Schr\"{o}dinger equation''
\begin{equation}
	\frac{\partial|\Phi(t)\rangle}{\partial
	t}=-H|\Phi(t)\rangle \Rightarrow|\Phi(t)\rangle=\exp
	(-H t)|\Phi(0)\rangle\,,
\label{eqs:pesudoH}
\end{equation}
where the pseudo-Hamiltonian $H$ is generally not Hermitian.

In terms of the ladder operator language, the gain and the loss terms 
originating from the master equation for $P\left(\left\{n,m\right\};t\right)$ 
are embedded in $H$. The rule of thumb is that the losses of particles give 
rise to the positive loss terms with the number operators $\ha_{i}a_i$ and
$\hb_{i}b_i$ being raised to the normal-ordered powers of corresponding 
reactant changes, and the negative terms for the gain balance directly 
reflect how many particles are destroyed and (re-)created.  For example, 
considering the reaction $k I\xrightarrow{\alpha} l I$ without
diffusion, one obtains $H_{\mathrm{react}}=\alpha\sum_{i}
\left(\ha_{i}^{k} - \ha_{i}^{l}\right) a_{i}^{k}$. Diffusion between 
neighboring sites $i$ and $j$ is nothing else but just the reactions $I_i
\xleftrightarrow{D_{I0}} I_j$ and $R_i\xleftrightarrow{D_{R0}} R_j$, with the
microscopic diffusion rates $D_{I0}$ and $D_{R0}$. Hence, the reaction 
scheme \eqref{eqs:allreact}, when supplemented with diffusions of both 
species, yields $H=H_{\mathrm{diff}}+ H_{\mathrm{react}}$ with
\begin{IEEEeqnarray}{rCl}
	H_{\mathrm{diff}} &=& \sum_{\langle i j\rangle}\Big[D_{I0}
	\left(\ha_{i}- \ha_{j}\right)\left(a_{i}-a_{j}\right) \nonumber\\
	&& \qquad + D_{R0}\left(\hb_{i}- \hb_{j}\right)\left(b_{i}-b_{j}\right)
        \Big] \,, \IEEEyesnumber* \IEEEyessubnumber*\\
	H_{\mathrm{react}} &=& \sum_{i}\Big[\lambda(1-\ha_i)\ha_i a_i
	+ \kappa (\ha_i-1)\ha_i a_i^2 + \sigma(\ha_i-1)a_i \nonumber \\
        && \qquad +\mu(\ha_i-\hb_i)a_i + \nu(\ha_i -1)\hb_i b_i
        a_i\Big]\,.
\label{eqs:hamilt}
\end{IEEEeqnarray}

The field theory action will take its shape within the exponential weight 
for the statistical average of an arbitrary observable $\mathcal{O}$. To 
this end, by introducing the projection state $\langle \mathcal{P}|=\langle
0|\prod_{i} e^{a_i+b_i}$, which satisfies $\langle \mathcal{P}|\ha_i=\langle
\mathcal{P}| = \langle \mathcal{P}| \hb_{i}$, the expectation value of
$\mathcal{O}$ reads
\begin{IEEEeqnarray}{rCl}
	\angs{\mathcal{O}(t)} &=& \sum_{\{n,m\}} \mathcal{O} (\{n,m\})
	P(\{n,m\};t) \nonumber \\&=&  \langle \mathcal{P}|\mathcal{O} 
	(\{\ha a,\hb b\}) |\Phi(t)\rangle \nonumber\\
	&=& \langle \mathcal{P}|\mathcal{O} 
	(\{\ha a,\hb b\}) \exp(-Ht)|\Phi(0)\rangle\,.
	\label{eqs:average}
\end{IEEEeqnarray}
Next, we follow the standard path integral
construction~\cite{tauber2005app,tauber2014} by splitting the 
temporal evolution $\exp(-Ht)$ into infinitesimal increments and
inserting at each time step the identity operator
$\mathbf{1}=\int\prod_{i}\left(\frac{\ud^2 \phi}{\pi}
\right)\left(\frac{\ud^2 \varphi}{\pi} \right) |\{\phi,\varphi\}\rangle
\langle \{\phi,\varphi\}|$, where the coherent states $|\phi_i\rangle$
and $|\varphi_i\rangle$ are right eigenstates of the the annihilation
operators, $a_i|\phi_i\rangle=\phi_i|\phi_i\rangle$ and
$b_i|\varphi_i\rangle=\varphi_i|\varphi_i\rangle$, permitting a
transformation from q-numbers ($\ha_i$, $a_i$, $\hb_i$, $b_i$) to
c-numbers ($\phi_i^*$, $\phi_i$, $\varphi_i^*$, $\varphi_i$). After 
further taking the continuum limit, $\sum_{i}\to h^{-d} \int \ud^d x$,
$\phi_i^{*}\to \ha(x,t)$, $\phi_i \to h^{-d} a(x,t)$, $\varphi_i^{*}\to
\hb(x,t)$, $\varphi_i \to h^{-d} b(x,t)$, where $h$ denotes the lattice
spacing, the resulting statistical average becomes
\begin{equation}
	\angs{\mathcal{O}(t)}=\int \mathcal{D}[\ha,\hb,a,b]
	\mathcal{O}(\{\ha a, \hb b\})\exp(-A[\ha,\hb, a, b])\,,
\end{equation}
with the field theory action
\begin{widetext}
\begin{equation}
	A=\int \ud^d x \ud t\Big[(\ha-1)\partial_t a + D_{I}'\nabla \ha
	     \cdot \nabla a + (\ha-1)(\sigma-\lambda \ha+\kappa' \ha a)
	     a +\left( \hb-1 \right)\partial_t b + D_{R}' \nabla \hb \cdot
     \nabla a + \mu(\ha - \hb)a+\nu' (\ha-1)\hb b a \Big]\,.
\label{eqs:actinit}
\end{equation}
\end{widetext}
Note that the microscopic diffusion rates have been replaced by the
continuum diffusivities $D_{I/R}'=h^2 D_{I0/R0}$, and the rates
$\kappa'=h^d \kappa$ and $\nu'=h^d \nu$. In the above expression, the
terms $\int \ud^dx \ud t\, \ha \partial_t a$ and $\int \ud^dx \ud t\,
\hb \partial_t b$ stemming from the initial and the final factors 
of the path integral have also been kept. As it is standard, the 
time limit in the action can be formally taken from $-\infty$ to
$\infty$.

The fields $\ha(x,t)$, $a(x,t)$, $\hb(x,t)$, and $b(x,t)$ in
\eqref{eqs:actinit} are still complex. In order to relate them to the
density fields $\rI(x,t)$ and $\rR(x,t)$, one can utilize the fact that 
$\ha(x,t)a(x,t)=\rI(x,t)=\exp(\trI)\rI\exp(-\trI)$ and $\hb(x,t)
b(x,t)=\rR(x,t)=\exp(\trR)\rR\exp(-\trR)$, with the auxiliary (imaginary)
response fields $\trI(x,t)$ and $\trR(x,t)$. Since $a(x,t)$ and $b(x,t)$
carry dimensions of particle densities after performing the continuum
limit, we can make the ansatz $\ha=\exp(\trI)$, $a=\rI\exp(-\trI)$,
$\hb=\exp(\trR)$, and $b=\rR\exp(-\trR)$ to construct a quasi-canonical
transformation \cite{janssen2005field}. Upon employing this
transformation, followed by the expansion of the exponentials,
integrating by parts, and discarding fourth- and higher-order terms,
the action finally takes the following form
\begin{widetext}
	\begin{IEEEeqnarray}{rCl}
	A&=&\int \ud^d x \ud t\Big\{\trI\partial_t \rI -
		\trI\rI\partial_t\trI  - D_{I}'\left[\trI \nabla^2 \rI +
		\rI(\nabla\trI)^2\right] + \trI\left[(\sigma+\mu-\lambda)
		+ \kappa'\rI + \nu'\rR - \frac{\lambda + \sigma}{2} \trI
	\right]\rI \nonumber \\
	&& \qquad \quad \,\,\,\,+ \trR\partial_t \rR -\trR\rR\partial_t\trR  - D_{R}'\left[\trR
		\nabla^2 \rR + \rR(\nabla\trR)^2\right] -\mu \trR
	\rI\Big\}\,.
\label{eqs:acttrans}
\end{IEEEeqnarray}
\end{widetext}
As one rescales the lengths, each term should acquire its respective
renormalized (running) coupling constant at different scales. Hence, in
the effective field theory, the above expression is rewritten 
in the following general way
\begin{widetext}
	\begin{IEEEeqnarray}{rCl}
	\mathcal{A}&=&\int \ud^d x \ud t\Big\{\trI\partial_t \rI -
		c \trI\rI\partial_t\trI  - D_{I1}\trI \nabla^2 \rI +
		D_{I2}\rI(\nabla\trI)^2 + \tau' \trI \rI + \trI\left[
		g_1 \rI + g_2 \rR - g_3 \trI \right]\rI \nonumber \\
	&& \qquad \quad \,\,\,\,+ \trR\partial_t \rR
	-c'\trR\rR\partial_t\trR  - D_{R1}\trR \nabla^2 \rR +
        D_{R2}\rR(\nabla\trR)^2 -g_4 \trR \rI\Big\}\,.
\label{eqs:actgen}
\end{IEEEeqnarray}
\end{widetext}

Following the same arguments in Ref.~\cite{janssen2005field}, the
coupling $g_1$ turns out to be irrelevant, and the fields as well as 
the couplings $g_2$ and $g_3$ are rescaled by a dimensionful amplitude
$K$ as $\tI=K^{-1}\trI$, $\sI=K\rI$, $\tR=K^{-1}\trR$, $\sR=K\rR$, and
$K g_2=2K^{-1}g_3=g'$, leading to
\begin{widetext}
	\begin{IEEEeqnarray}{rCl}
	\mathcal{A}&=&\int \ud^d x \ud t\Big\{\tI\partial_t \sI -
		c/K \tI\sI\partial_t\tI  - D_{I1}\tI \nabla^2 \sI +
		D_{I2}/K \sI(\nabla\tI)^2 + \tau' \tI \sI +
		\frac{g'}{2}\tI\left(
		2\sR - \tI \right)\sI \nonumber \\
	&& \qquad \quad \,\,\,\,+ \tR\partial_t \sR
	-c'/K\tR\sR\partial_t\tR  - D_{R1}\tR \nabla^2 \sR +
       D_{R2}/K \sR(\nabla\tR)^2 -g_4 \tR \sI\Big\}\,,
\label{eqs:actgen1}
\end{IEEEeqnarray}
\end{widetext}
in which, upon rescaling the spatial distances $x$ by a length scale 
$\kappa^{-1}$, the naive scaling dimensions of the fields and couplings are
fixed to $[\sR]_0\sim [\tilde{\sI}]_0 \sim \kappa^{(d-2)/2}$, $[\sI]_0
\sim [\tilde{\sR}]_0 \sim \kappa^{(d+2)/2}$, $[\tau]_0\sim \kappa^{2}$,
$\quad [g]_0\sim \kappa^{(6-d)/2}$, $[g_4]_0\sim \kappa^0$, $[c/K]_0\sim
[D_{I2}/K]_0 \sim \kappa^{-(d-2)/2}$, $[c'/K]_0\sim [D_{R2}/K]_0
\sim \kappa^{-(d+2)/2}$. Consequently, the upper critical dimension $d_c=6$ 
remains the same as the DIP, while the couplings $c/K$, $c'/K$,
$D_{I2}/K$, and $D_{R2}/K$ are all irrelevant near $d_c$, rendering
	\begin{IEEEeqnarray}{rCl}
	\mathcal{A}&=&\!\!\int\!\! \ud^d x \ud t\Big\{\tI\partial_t \sI - D_{I1}\tI
		\nabla^2 \sI + \tau' \tI \sI +
		\frac{g'}{2}\tI\big(
		2\sR - \tI \big)\sI\nonumber \\
	&& \qquad \quad   + \tR\partial_t \sR - D_{R1}\tR \nabla^2
        \sR - g_4 \tR \sI\Big\}\,,
\label{eqs:actgen2}
\end{IEEEeqnarray}

Finally, we note that $g_4$ is naively dimensionless and what is more, 
there are no diagrams to renormalize it, implying that it will remain
dimensionless with respect to rescaling. Hence, it is customary to rescale 
the time $t\to t'=g_4 t$ to eliminate $g_4$. Upon renaming the couplings 
to $D_{I}=D_{I1}/g_4$, $D_{R}=D_{R1}/g_4$, $D\tau=\tau'/g_4$, and
$g=g'/g_4$, we arrive at the action \eqref{eqs:action}.
%

\end{document}